\documentstyle[12pt]{article} \textwidth= 15cm \textheight= 20.5cm
\begin{document}
\begin{flushright} THEP 93/24\\ Univ. Freiburg\\ hep-th/9310037\\
September 1993\\ \end{flushright}
\begin{center} \vskip 0.5cm
{\Large  Gauging and Symplectic Blowing Up in Nonlinear }\\
\vskip 0.4cm
{\Large Sigma Models: I. Point Singularities}\\
 \vskip 1.5cm
H. B. Gao\footnote{Alexander von Humboldt fellow, on leave  from Zhejiang
University, Institute of Modern Physics, Hangzhou, China}
{\it and} H. R\"omer\\
\vskip 0.5cm
{\it Fakult\"at f\"ur Physik\\
Universit\"at Freiburg\\ Hermann-Herder-Str. 3\\ D-79104 Freiburg, Germany}\\
\vskip 1.5cm
{\sc Abstract}\\
\end{center}

In this paper a two dimensional non-linear sigma model with a general
symplectic manifold with isometry as target space
is used to study symplectic blowing
up of a point singularity on the zero level set of the moment map associated
with a quasi-free Hamiltonian action. We discuss in general the relation
between symplectic reduction and gauging of the symplectic isometries of
the sigma model action. In the case of singular reduction, gauging has
the same
effect as blowing up the singular point by a small amount. Using the
 exponential mapping
of the underlying metric, we are able to construct symplectic
diffeomorphisms needed to glue the blow-up to the global reduced space
which is regular,
thus providing a transition from one symplectic sigma model to another one
free of singularities.

\newpage
\noindent {\bf I. Introduction}

A general nonlinear sigma model with  some Riemannian target manifold $M$
admits as a global
symmetry the isometry group of $M$, which can be gauged [1]
by introducing a minimal coupling to a gauge field. The same can be done in
the presence of more structures on $M$, e.g. for the supersymmetric nonlinear
sigma models [2]-[3].
In addition to its application to SUSY phenomenology, an important use of
the sigma model gauging is to construct the quotient space and define on
it a reduced nonlinear $\sigma$-model
with fewer degrees of freedom ( see [4] for a review of various quotient
constructions including hyperk\"ahler manifolds ). One recent example is
the gauging of a WZW model which
leads to a quotient space describing the black hole geometry [5].

The nonlinear $\sigma$-model obtained by gauging can be very different
from the original
one, eventhough it is possible to return to the original number of degrees
of  freedom by introducing Lagrangian
multiplier fields. Especially if the isometry group of the starting
nonlinear $\sigma$-model
has a nontrivial isotropy subgroup at certain points, the resulting
quotient space in general
contains  singularity points. A less trivial construction [6] (which
is related to the duality transformation) reveals further that there
are equivalent gaugings in a single model, one of which leads to a singular
quotient while another
smooth. This raises the interesting question of whether a smooth change of
topological structures in spacetime can be realized
 within a higher symmetric (string) theory [7].

In this paper, we give a detailed construction of gauging a quasi-free
group action in a general nonlinear $\sigma$-model on a symplectic manifold.
Usually, minimal
coupling is not enough for gauging a general $\sigma$-model with a WZ
term [8], but in the
presence of an almost complex and symplectic structure which we assume,
it is still possible to
use minimal coupling to the gauge field. The nontrivial point comes when
 dealing with the quasi-free group action. In that case, the symplectic
quotient
is not necessarily smooth: it may contain point-like singularities. The
singularities are precisely points
where the isotropy subgroups are nontrivial. There are many reasons to
believe
that the physics near the singularity is far from trivial. In fact orbifolds
(those singular spaces whose singularities have only discrete isotropy groups)
in both quantum mechanics [9] and string theory [10] have been invoked as
a major probe of higher (quantum) symmetries of the theories. Although
singularities in e.g. stringy orbifolds look harmless due to the existence of
winding modes, many
theories lack efficient calculability when dealing with singularities. The
question then arises  whether we are able to make some predictions
regarding behaviors of the theory near the singularities by looking at the
blow-up of the singular space. Since blowing up is in general a mild
operation
(it is even a bi-rational transformation) it may turn out that one obtains
equivalent theories by going over to the blow-ups. However, unlike the
previous examples of duality transformations in conformal field theory which
can be taken as certain global discrete symmetries, at least at the tree
level, blowing up operations usually involve some
diffeomorphisms preserving the K\"ahler or symplectic structures on the
manifold, and therefore
must be a symmetry of only diffeomorphism-invariant theory.
To find out this diffeomorphism-invariant theory is of course of
utmost interests.

Blowing up a singularity in the complex category
is a more or less familiar procedure [11].
Some of its rudiments with application to construct
mirror pairs of Calabi-Yau spaces can be found in ref [12]. See also [13]
for a use
of blowing up orbifold points in constructing a geometric realization of the
conjectured equivalence between Landau-Ginsburg and Calabi-Yau descriptions
of the string backgrounds.
For blowing up in the
symplectic category, however, the relevant notions appear only recently
[14]-[16]. Let us explain roughly what is involved in this construction.
Since locally all symplectic manifolds look alike, symplectic invariants
of a
typical symplectic manifold are global in nature. Thus when we calculate for
example the cohomology classes $[\omega_t]\in H^2(M)$ for a family of
symplectic
2-forms parametrized by a real variable $t$, the result does not vary much
as we go along a smooth trajectory in $M$ with the parameter $t$.
In fact, according to the Duistermaat-Heckman theorem [17], $[\omega_t]$
traces
a  line in the space $H^2(M)$  if the trajectory of $t$ does not cross
 a singular point. It  is especially interesting when  we think of the
symplectic manifolds parametrized by $t$ as a sequence of Marsden-Weinstein
reduced symplectic spaces, then $t$ is provided by (regular) values of the
moment
map sitting in a region in the dual space of the Lie algebra of the
Hamiltonian vectors. If the corresponding group
 is abelian, e.g., a torus $T^n$, by the well known theorem of
Atiyah and Guillemin-Sternberg [18], the image of the symplectic manifold
under the $T^n$-equivariant moment map is a convex body  (polytope for
example).
In this case the regular values of the moment map at which the
Marsden-Weinstein
reduction is performed lie all in the interior region. When one tries to
pass over the borderline of two separate subpolytopes the reduced space
becomes singular. Thus a sequence of M-W reductions can be used to probe the
local structures near the singular points. Now the M-W reduction for the local
model of a symplectic  manifold is extremely simple, in fact even a single
generating symplectic space for a variety of reduced spaces can be
constructed [15]
which is much like the generalization of  differential topological cobordism.
Locally a symplectic blowing up contains nothing more than the same operation
in complex analytic terms, i.e., replacing the linar space by a projective
space (or more precisely a line bundle over the projective space made up by
incidence relations). Globally, however, several symplectic diffeomorphisms
are needed to
glue the blow-ups smoothly back to the complement  of the singular points of
the original symplectic manifold. The resulting smooth manifold possesses a
well defined reduced symplectic form, making it a genuine symplectic manifold
which may differ from the original one by change of topological properties,
for example.

The main concern of this work is to provide a gauged nonlinear sigma model
version
of the above constructions. In a subsequent paper, we will generalize this
construction to the case of blowing up along a singular submanifold.
The following is a brief outline of this paper.
In section II, we collect the basic ingredients of nonlinear sigma model
with a general symplectic target manifold $M$, suitable for gauging of the
isometries
preserving symplectic structure. In addition to the standard action, we also
include a linear sigma model action in terms of normal coordinates centered
at the specific point on the manifold which is a fixed point of a
Hamiltonian
group action of quasi-free type. We then discuss in section III in general
terms the gauging of
symplectic isometries leading to the symplectic quotient. This involves
spelling
out explicitly the moment map constraint in terms of a suitable coordinate
system defined by solutions of the Pfaff equation associated to an integrable
distribution. Gauging thus can be implemented as in the usual topological
quotient
construction on the zero level set of the moment map.
In section IV, we deal with the quasi-free $U(1)$ action on the
linear sigma model whose fields take values in a small neighborhood of
the singular point
which is diffeomorphic to $C^n$, we show the exact relationship between
blowing up the
singular point and the gauging of the corresponding group action. In
section V,
we discuss the relation of this linear model to the nonlinear one,
emphasizing the importance of symplectic diffeomorphisms. We use the normal
coordinate expansion
method to derive a (weakly) coupled form of the total action, and argue that
the decoupling is implemented by the one loop effective action of the
original nonlinear sigma model. We also discuss how to compare gauge fields
arising from gauging both linear and nonlinear $\sigma$-models.
Section VI contains two examples of the application of our general results.
Section VII summarizes our conclusions.
An appendix contains a proof of the existence of the change of coordinates
used in section III.
\vskip 1cm
\noindent {\bf II. The Lagrangian}

Our starting point is a nonlinear $\sigma$-model, with  an arbitrary
almost complex, symplectic target manifold
$M$, over a Riemann surface $\Sigma$. Thus one introduces
a complex structure on the tangent space of $M$ (dim $M$=m ) with
tensor field
$F^i_j$ satisfying
$$ F^i_kF^k_j=-\delta^i_j, ~~~~ i.j=1,... m. \eqno(2.1)$$
This gives an almost complex structure to $M$.
A symplectic structure on $M$ is provided if there exits a (globally defined)
closed 2-form $\omega$:
$$ d\omega=0,$$
$$\omega=F_{ij} dx^i\wedge dx^j, \eqno(2.2)$$
$$ F_{ij}=G_{ik}F^k_j.$$
where $G_{ij}$ is a Riemannian metric on $M$ compatible with the almost
complex structure $F^i_j$, i.e.
$$G_{ij}=F^k_iF^l_jG_{kl}. \eqno(2.3)$$
The last property is equivalent to the antisymmetry of $F_{ij},
F_{ij}=-F_{ji}$.
(A calibrated almost complex structure.)
Let  $\epsilon^{\alpha}_{\beta}$ be the natural complex structure on the
Riemann
surface $\Sigma$, and $\epsilon_{\mu\nu}=h_{\mu\alpha}\epsilon^{\alpha}_{\nu}$,
$\epsilon_{\mu\nu}=-\epsilon_{\nu\mu}$ be the antisymmetric rank-2 tensor on
$\Sigma$ with
metric $h_{\mu\nu}$. We adopt the notation where the integral of the 2-form
$\omega$ is expressed in terms of the $\sigma$-model scalar fields, i.e.
$$\int F_{ij}dx^i\wedge dx^j =\int d^2 \sigma\epsilon^{\mu\nu}F_{ij}
\partial_{\mu}x^i \partial_{\nu}x^j. \eqno(2.4)$$

Let the manifold $M$ be parametrized by $m$ coordinate scalars $\Phi^i$.
Our general
symplectic nonlinear $\sigma$-model has the following action (here, as usual,
we have
suppressed the 2d metric $h_{\mu\nu}$, in order to avoid notational
complexity):
$$S={1\over2}\int d^2\sigma G_{ij}\partial^{\mu}\Phi^i\partial_{\mu}\Phi^j
+{1\over2}\int d^2\sigma \epsilon^{\mu\nu}F_{ij}\partial_{\mu}\Phi^i
\partial_{\nu}\Phi^j,
\eqno(2.5)$$
 with $F_{ij}$ satisfying eq(2.2). Since the second term in eq(2.5) is
topological
the invariance of the action (2.5) is the same as that of the first term, i.e.
the whole isometry group of $M$ (as a riemannian manifold). However, when
 gauging
is concerned, not all isometries can be gauged, but only those preserving
the symplectic
form (2.4). The condition for an isometry
to preserve the symplectic form is familiar,
$${\cal L}_{\xi}\omega=di(\xi)\omega=0, \eqno(2.6)$$
where $\xi^i_a$ is a Killing vector generating the isometry (global)
transformation
$$\delta\Phi^i=\lambda^a\xi^i_a(\Phi). \eqno(2.7)$$
$\xi^i_a$ being a Killing vector means that the following Killing's equation
is satisfied
$$\nabla_{(i}\xi_{j)a}=0, \eqno(2.8)$$
where $\nabla_i$ denotes the covariant derivative with respect to the unique
Riemannian connection
$\Gamma^i_{jk}$ compatible with the metric $G_{ij}$ on $M$.

In the most general situation, we will assume that
the Killing vectors generate a group $K$,
$$[\xi_a,\xi_b]= {\cal L}_a\xi_b=f^c_{ab}\xi_c \eqno(2.9)$$
with $f^a_{bc}$ the structure constants of $K$.

Isometries fulfilling conditions (2.6) and (2.8) will be called symplectic
isometries.
We will consider a subgroup $H\subset K$ of the symplectic isometries
which leave
one point $x$ in $M$ fixed. The isotropy group at $x, K_x$ is the whole
of $H$ and we will gauge the subgroup $H$ by minimal coupling
to some gauge fields. In section III, we will argue that minimal coupling is
appropriate in this situation and also describe in some details the main
points
which arise in obtaining the symplectic quotient by gauging the symplectic
isometries.

In general, an isometry acts on $\Phi^i$ nonlinearly, while at
each point of the manifold $M$, local coordinates can be introduced such that
the isoptropy group at that point acts linearly (i.e, as in flat case).
This is
related to the fact that there always exists a normal coordinate neighborhood
of a given
point which is diffeomorphic to a small neighborhood of the origin in the
tangent space at that point [19]. The diffeomorphism in question is the
exponential mapping defined by the underlying Riemannian structure (metric) of
$M$. We will say more on this diffeomorphism in section V. Here we content
ourselves by choosing a normal coordinate system $\{\phi^i, i=1,...,m\}$ in an
$\epsilon$-small neighborhood $U_{\epsilon}$ of $x\in M$, in terms of which
we have a free, linear $\sigma$-model in addition to the one given by (2.5),
$$S_0={1\over2}\int d^2x\partial^{\mu}\phi^i\partial_{\mu}\phi^j.
\eqno(2.10)$$

We remark that, when treated as  high energy modes of our nonlinear model
this linear
$\sigma$-model action can be conveniently integrated out resulting in some
world sheet effect which might in the present context be completly ignored.
However, we will not pursue this line of reasoning in this paper, and
instead
 treat this linear model completely on equal footing. This is in fact a
basic practice in symplectic dynamics, and we will directly deal with this
linear model when performing symplectic blowing up in section IV.
\vskip 1cm
\newpage
\noindent {\bf III. Symplectic quotient and gauging}

In this section, we describe in some details the gauged $\sigma$-model
realization of the regular symplectic reduction, which serves as the reduced
space outside the singular locus. The methods are quite general and may
have their own applications.

According to the general philosophy of $\sigma$-model isometry gauging, the
global invariance of eq(2.5) and eq(2.7) can be promoted to a local invariance
(with $\lambda^a(x)$ arbitrary functions of $x$) by introducing a gauge field
$A^a_{\mu}$ transforming as
$$\delta A^a_{\mu}=\partial_{\mu}\lambda^a+f^a_{bc}A^a_{\mu}\lambda^c.
\eqno(3.1)$$
Under  the transformation (3.1) and (2.7), the
 gauge covariant derivative defined by
$$D_{\mu}\Phi^i=\partial_{\mu}\Phi^i-A^a_{\mu}\xi^i_a(\Phi) \eqno(3.2)$$
transforms like a tangent vector,
$$\delta D_{\mu}\Phi^i=\lambda^a\partial\xi^i_a/\partial\Phi^kD_{\mu}\Phi^k.
\eqno(3.3)$$
This, together with eq(2.8) and the closedness of the symplectic 2-form,
guarantee the local gauge invariance of the minimally coupled, gauged action
$$S_{gauged}={1\over2}\int d^2xG_{ij}D^\mu\Phi^iD_\mu\Phi^j
+{1\over2}\int d^2x \epsilon^{\mu\nu}F_{ij}D_\mu\Phi^iD_\nu\Phi^j.
\eqno(3.4)$$
(Note that after the substitution of covariant derivatives, the second term
in (3.4) is no longer topological, but the reduced form after integrating out
the gauge fields will still be topological.)
As usual, we do not include kinetic terms for the gauge fields, so that the
gauged
model eq(3.4), after eliminating gauge fields, describes a nonlinear
$\sigma$-model
on the space of orbits of the group action on $M$. This is the usual geometric
interpretation of gauging a (Riemannian) isometry.

Gauging of symplectic isometries, leading to a symplectic quotient, has to
meet
additional constraints, i.e. eq(2.6). Locally (2.6) implies the existence of
a function $\mu$ such that
$$\omega_{ij}\xi^j_a =\partial_i\mu_a. \eqno(3.5)$$
Globally, the solution to eq(3.5) may not always exist. But under certain
assumptions, such as triviality of the first cohomology group of
$h (=Lie(H))$, global functions $\mu_a$ exist and are unique up to central
elements of $h$ [20]. These $\mu_a$ fit together to form the moment map
$$\mu: ~M \longrightarrow ~h^*. \eqno(3.6)$$
The symplectic quotient is the usual topological quotient of the subspace
$N=\mu^{-1}(0)$
by the group $H$. In our case, we have to implement the constraint
$\mu(\Phi)=0$
to the gauged action so that it really describes $N=\mu^{-1}(0)\subset M$.

Explicitly solving  the moment map constraints is in general a difficult
task, especially when the hamiltonian group action is nonlinearly realized.
However, when the condition on cohomology groups of the Lie algebra, e.g.,
$H^1(h)=H^2(h)=0$ is satisfied, one can convince oneself quickly that
components
of the moment map can be found in terms of Killing vectors. Indeed, since
$\xi_a \in h$ preserve the symplectic form $\omega=F_{ij}d\Phi^i\wedge
d\Phi^j$,
$${\cal L}_a(i(\xi_b)\omega)=i({\cal L}_a\xi_b)\omega+i(\xi_b)
{\cal L}_a\omega$$
$$=i([\xi_a,\xi_b])\omega=f^c_{ab} i(\xi_c)\omega. \eqno(3.7)$$
Thus the moment map components corresponding to Killing vectors $\xi_a$
can be written as
$$\mu_a=(f^a_{bc})^{-1} i(\xi_b)i(\xi_c)\omega=(f^a_{bc})^{-1}\xi^i_b
\xi^j_c\omega_{ij}. \eqno(3.8)$$
Note that $\mu_a$ in (3.8) is well defined if the Lie algebra is semi-simple,
that is if $f$ is invertible. So we have in principle $dimH$ relations
 expressed by
the vanishing of the components of the moment map (3.8). Although
reasonable,
it is in practice impossible to impose these constraints directly into the
Lagrangian, because of lack knowledge of the explicit dependence
of the Killing vectors on the coordinate
fields $\Phi^i$. We will devote the following  paragraphes to construct
a different
way to impose the moment map constraints directly into the Lagrangian.

Let us denote by $N$ the constraint set $\mu^{-1}(0)$ in $M$,
i.e. the set of points in $M$ which are mapped into the same point
$0\in h^*$.
To describe the submanifold $N=\mu^{-1}(0)$, first observe that because of
the equivariance of the moment map $\mu$,
$$\mu(g\cdot x)=g\cdot \mu(x) ~\in {\cal O} \subset h^*, \eqno(3.9)$$
for any point $x\in M$. Here $\cal O$ is an orbit in $h^*$ by the coadjoint
action. If $x$ lies in $N$, we must have $\mu(g\cdot x)=g\cdot \mu(x)=0$
for all
$x\in N$, since the $H$ action certainly leaves $0\in h^*$ fixed. Consider
the tangent space to the orbit $\cal O$ at any point $a\in {\cal O},
T_a{\cal O}$,
the lift to the tangent space $T_xM$ of the moment map $d\mu_x :
T_xM\rightarrow
h^*$ is defined as the transpose of the linear map $h\rightarrow T_xM$
(with $T_xM$ identified with $T^*_xM $ via the symplectic structure), sending
each element $\xi\in h$ into the Hamiltonian (tangent) vectors $\xi_M$ on $M$.
If we set
$$T_xN'=d\mu^{-1}_x(T_{\mu(x)=a}{\cal O}), \eqno(3.10)$$
as a submanifold of $T_xM$, it must be spanned by the basis of $T_a{\cal O}$
and those of $Ker d\mu_x$. Now $Ker d\mu_x$ can be taken as a definition for
a vector $Y$ to be an element of $T_xN$
$$d\mu_x(Y)=0 ~~~ \forall Y \in T_xN. \eqno(3.11)$$
Thus the basis vectors in $T_a{\cal O}$ are in one to one correspondence
 with the
vectors lying in the orthogonal complement of $T_xN$ in $T_xM$. Obviously,
the complement of the tangent vectors coming from $h$ in $T_xM$ is contained
in $T_xN$.

Now the condition for a vector $Y$ to be tangent to $N$ at $x$ can be written
as
$$i(\xi)\omega(Y)=d\mu(Y)=0, ~~\forall \xi \in h .\eqno(3.12)$$
Because $\xi$ is a symplectic vector field, i.e. $d i(\xi)\omega=0$, the above
equation (3.12) defines an integrable distribution
or, in  slightly different terms, a
foliation [20]. The equivariance of the moment map guarantees that this
foliation
is in fact fibrating, i.e., in a certain coordinate system, the solution
(integration) of the equation (3.12) defines a submanifold $N$ in $M$. The
desired
coordinate system is provided by the Frobenius theorem [20], which says that
there exists a coordinate system $ \{ w^i, i=1,...,m\}$ in the neighborhood
$\cal W$  of each point in $M$, so that the leaves of the
foliation are given by
$$w^1=const.,~ .~.~.~, w^r=const., ~~~r=m-dimH=dim N, \eqno(3.13)$$
and that the tangent space at that point to the foliation is spanned by
$$ {\partial\over{\partial w^{r+1}}},~.~.~.,~ {\partial\over{\partial w^m}}.
\eqno(3.14)$$
In this case the (co)tangent space of the submanifold $N\subset M$ is the null
subspace of $T^*_m M$ with respect to the distribution.
In the coordinate system $\{ w^i\}$ the symplectic form is
$\omega'=\sum {\omega'}_{ij}dw^i \wedge dw^j$, with the functions
 ${\omega'}_{ij}$ satisfying
$${\omega'}_{ij}=0,~~~i,j= r+1,...,m, \eqno(3.15)$$
$${{\partial {\omega'}_{ij}}\over{\partial w^a}}=0,~~~a=r+1,...,m.
\eqno(3.16)$$
The first is due to the condition (3.12), while the second is due to
$d\omega=0$.
If the foliation is fibrating (which is the case here), then
$w^s, s=1,..., r$ can be chosen as the coordinates of $N=\mu^{-1}(0)$ on
which there exists a closed 2-form $\omega^*$ with
$f^*\omega'=\omega^*$ where $f: (w^1,... w^m) \rightarrow (w^1,... w^r)$
is a submersion. Now, the question boils down to find the system of
coordinates dictated by the Frobenius theorem.

In the above discussion, we have used the fact that the set of
Hamiltonian vectors $\xi_a$ are identified with
(smooth) tangent vectors to $M$, with
respect to an arbitrary
coordinate system. In order to apply the Frobenius theorem, $\xi_a$ have
to be converted into the form (3.14) by a suitable change of coordinates.
Note that in an arbitrary
coordinate system $({\cal U}; u^i)$, $\xi_a$ looks as
$$\xi_a\vert_{\cal U}=\sum \xi^i_a {\partial\over{\partial u^i}}.
\eqno(3.17)$$
Comparing with (3.14), we see that we must find a change of coordinates which
smears out a sufficient number of components of $\xi_a$. In fact, if there
exists a change
of coordinates, $u^i \rightarrow v^i$, such that, in the coordiante system
$({\cal V};v^i)$, $\xi_a$ looks like
$$\xi_a\vert_{\cal V}=\xi^1_a(v){\partial\over{\partial v^1}} \eqno(3.18)$$
for each $a$, then, by defining $w^1=\int^{v^1}_0 dv^1/\xi^1, ~
 w^{\alpha}=v^{\alpha}$,
$\alpha=2,..., m$, we have
$$\xi_a\vert_{\cal W}={\partial\over{\partial w^1}} \eqno(3.19)$$
for each $a$. Continuing this step for every $\xi_a, a=r+1,..., m$, we thus
have provided the desired system of coordinates $({\cal W}; w^i)$ in terms of
which the Hamiltonian vectors $\xi_a$ are converted into the form (3.14).
In Appendix A, we prove that there indeed exists such a change of coordinates
with the above properties. Now in the new coordinate system, we are able to
 write down
the gauged action directly in terms of coordinates $w^i$ on $N=\mu^{-1}(0)$.
Using equations (3.15), (3.16) together with
$$F'_{ij}=G'_{ik}F^k_j, \eqno(3.20)$$
it is straightforward to verify that the resulting gauged action, in the new
coordinate system, takes the
same form as in (3.4) except now the indices $i,j$ run from $1,..., dimN$,
 and the functions $G_{ij}, F_{ij}$ become
$$G'_{ij}=G_{\alpha\beta}{{\partial w^{\alpha}_{(0)}}\over{\partial
w^i_{(a)}}}
{{\partial w^{\beta}_{(0)}}\over{\partial w^j_{(a)}}}=G_{\alpha\beta}
{{\partial w^{\alpha}_{(0)}}\over{\partial w^{i_1}_{(1)}}} {{\partial
w^{i_1}_{(1)}}
\over{\partial w^{i_2}_{(2)}}}~.~~.~~.~~
{{\partial w^{i_{a-1}}_{(a-1)}}
\over{\partial w^i_{(a)}}} \times $$
$$\times {{\partial w^{\beta}_{(0)}}\over{\partial w^{j_1}_{(1)}}}
{{\partial w^{j_1}_{(1)}}\over{\partial w^{j_2}_{(2)}}}~.~~.~~.~~
{{\partial w^{j_{a-1}}_{(a-1)}}\over{\partial w^j_{(a)}}}, \eqno(3.21)$$

$$F'_{ij}=-G'_{i, j+{m\over 2}}, ~~ j \leq m/2,~~~~~
F'_{ij}=G'_{i, j-{m\over 2}}, ~~ j> m/2,
   \eqno(3.22)$$
where the partial derivatives are given by the relations
$${{\partial w^s_{(s-1)}}\over{\partial w^s_{(s)}}}=\xi^s_{(s)},~~~
{{\partial w^{\alpha}_{(s-1)}}\over{\partial w^s_{(s)}}}=\xi^{\alpha}_{(s)},
{}~~~
{{\partial w^{\alpha}_{(s-1)}}\over{\partial w^{\beta}_{(s)}}}=
\delta^{\alpha}_{\beta},$$
$$s=1,~.~.~.,~dim H;~~~\alpha, \beta=s+1, . . ., m. \eqno(3.23)$$

We  remark that after the  coordinate transformation, the Killing vectors
are now represented by a set of null vectors in $dimH$ directions on the
submanifold $N$:
$$\delta \Phi^s_{(a)}=\lambda^s ,~~ s=1,..., dim H. \eqno(3.24)$$
In the case of an abelian group $U(1)$, the above formulae simplify to
$$G'_{11}=G_{11}(\xi^1)^2+G_{1a}\xi^1\xi^a+G_{ab}\xi^a\xi^b,$$
$$G'_{a1}=G_{a1}\xi^1+G_{ab}\xi^b,$$
$$G'_{ab}=G_{ab}, ~~a, b\not= 1, 1+m/2 . \eqno(3.25)$$
and similarly for $F'_{ij} (F'_{aa}=0).$

If we denote
$$(\xi_m,\xi_n)=G_{ij}\xi^i_m\xi^j_n=G'_{mn},$$
$$\widehat{(\xi_m,\xi_n)}=F_{ij}\xi^i_m\xi^j_n=F'_{mn},$$
$$\xi_{ma}=G_{ia}\xi^i_m,~~~~ \widehat{\xi_{ma}}=F_{ia}\xi^i_m, \eqno(3.26)$$
where $m,n=1,..., dimH$, and $a,b=1,..., m-2dimH$, the general
gauged action takes the form
$$S_{gauged}={1\over 2}\int d^2\sigma [(\xi_m,\xi_n)D_{\mu}\Phi^m D^{\mu}
\Phi^n+\xi_{ma}D_{\mu}\Phi^m \partial^{\mu}\Phi^a]$$
$$+{1\over 2}\int d^2\sigma [\epsilon^{\mu\nu}\widehat{(\xi_m,\xi_n)}D_{\mu}
\Phi^m D_{\nu}\Phi^n+\epsilon^{\mu\nu}\widehat{\xi_{ma}}D_{\mu}\Phi^m
\partial_{\nu}\Phi^a]$$
$$+S[G_{ab}, F_{ab}],
\eqno(3.27)$$
$$D_{\mu}\Phi^m =\partial_{\mu}\Phi^m -A^m_{\mu}.$$
Solving the equation of motion for the nondynamical gauge fields
$$[\eta^{\mu\nu}(\xi_m,\xi_n) +\epsilon^{\mu\nu}\widehat{(\xi_m,\xi_n)}]
A^n_{\nu}
=\eta^{\mu\nu}(\xi_m,\xi_n)\partial_{\nu}\Phi^n$$
$$+\epsilon^{\mu\nu}\widehat{(\xi_m,\xi_n)}\partial_\nu \Phi^n
+{1\over 2}\eta^{\mu\nu}\xi_{ma}\partial_{\nu}\Phi^a +{1\over 2}
\epsilon^{\mu\nu}\widehat{\xi_{ma}}
\partial_{\nu}\Phi^a, \eqno(3.28)$$
and choosing a gauge, leads us back to a symplectic $\sigma$-model
defined
on the symplectic quotient space. It is easily checked that the solution of
(3.28) transforms as a connection of  the principal $H$ bundle.

In view of (3.24), what we have done above amounts to converting the general
$H$ action of isometries on $M$ into the Abelian torus (taking into account
 the periodic boundary conditions for $\Phi^m$) action on $N$, with
simplified gauge connections. This is important in order to compare with
the quotient in the linear case, to be considered in next section, which
arises from gauging only the Abelian subgroup of the whole $U(n)$ isometries.

\vskip 1cm
\noindent {\bf IV. The local model}

In a neighborhood of each point on a symplectic manifold, the symplectic
form can be brought into the standard form and the Hamiltonian group action
is linearly realized. Gauging thus becomes extremely simple. We will
consider in this
section the consequences of gauging a quasi-free $S^1$ action in a small
neighborhood of a point (a singular point of the group action, to be
precise)
where we have a linear $\sigma$-model given by (2.10). The origin of this
linear $\sigma$-model may be looked at from several viewpoints,
physically, it can be regarded
as the result of performing a normal coordinate expansion around a certain
classical configuration, see Section 5 for more. We only consider the
abelian $U(1)$ action, since this case is typical and sufficient to
 demonstrate the essence of symplectic blowing up.

An almost complex, symplectic manifold has  on its tangent space at a point
a natural complex structure. Thus we can use complex coordinates,
$z^i=1/{\sqrt 2}
(x^i+iy^i), ~ {\bar z}^i=1/{\sqrt 2}(x^i-iy^i)$, where $x, y$ are real fields
and the index runs now $ i=1,..., n=m/2$. In terms of the complex fields $z^i,
{\bar z}^i$, our linear $\sigma$-model action takes the form
$$L=\sum^n_{i=1}\int d^2 \sigma \partial_{\mu}z^i\partial^{\mu}{\bar z}^i
=\sum \int d^2 \sigma \partial_+ z^i \partial_-{\bar z}^i. \eqno(4.1)$$
The complex fields $ z^i, {\bar z}^i$ span a linear space $C^n$ on which
$U(1)$ acts linearly and analytically, preserving the symplectic structure
$$\Omega=-i \partial {\bar \partial} \vert z\vert^2=-i\sum dz^i \wedge
d{\bar z}^i. \eqno(4.2)$$
(Actually, the maximal analytic invariance of (4.2) is $U(n)$.)
Our assumption of the quasi-free action implies in this case the existence
of an isolated singularity. Without loss of generality, we can take the
global $U(1)$ action acting diagonally on $z^i$, as follows
$$z^1\rightarrow e^{-i\theta}z^1, ~~~~ z^i\rightarrow e^{i\theta}z^i, ~~~~
i=2,..., n, \eqno(4.3)$$
with $\theta \in {\bf R}$ constant.

The moment map associated to this action is
$$\mu(z)=-\vert z^1 \vert^2+\sum^n_{i=2}\vert z^i\vert^2. \eqno(4.4)$$
It is clear that $\mu^{-1}(0)$ contains an isolated critical point at $0\in
C^n$. Because of this singularity, pathology is to be envisaged when one
tries to perform the symplectic reduction. However, a genuine blow up makes
sense and relates the strata of the singular level set to some
 Marsden-Weinstein reduced spaces.

First we need to recall some facts concerning the mathematical notion of
blowing up \footnote{The following
 few paragraphs are probably well known to experts. The reader can find
most of the mathematical statements in ref[15][16], or consult [22] for a
similar discussion including some physical motivations}.
Suppose the origin $0$ in $C^{n+1}$ is a singular point, to be blown up.
Due to the
well known relation in complex geometry between $C^{n+1}-\{0\}$ and $CP^n$
[21], there is a line bundle $L$ over $CP^n$ whose fibers are copies
of $C^*=C-\{0\}$.
The blowing up of $C^{n+1}$ at the origin amounts to a map which sends the
complement of the zero section of $L$ bijectively onto the complement of the
origin in $C^{n+1}$, and sends the whole zero section to the complex
projective space $CP^n$. Then $L$ is called the blow up of $C^{n+1}$
at origin.

In the situation of (4.4), if we slightly perturb the value of the moment map
by $\pm\epsilon$ with $\epsilon > 0$ sufficiently small, and consider the
 Marsden-Weinstein
reduced spaces at both $+\epsilon$ and $-\epsilon$, we obtain two different
reduced spaces. On $-\epsilon$, it is a copy of $C^{n-1}$ through
identification
of the global cross section of the $U(1)$ action. On $+\epsilon$, it is the
line bundle $L$ over $CP^{n-2}$. In fact, for $\mu=+\epsilon$, the level set
 of the  moment map is
$$-\vert z^1\vert^2+\sum\vert z^i\vert^2=\epsilon. \eqno(4.5)$$
It becomes $S^{2n-3}\times C$ after a change of coordinates
$$w^1=z^1, ~~~ w^i=(\epsilon+\vert z^1\vert^2)^{-1/2} z^i. \eqno(4.6)$$
The $S^1$ action in terms of the new coordinates $w^1, w^i$ sends $w^1$ into
$e^{-i\theta'}w^1$ and $w^i$ into $ e^{i\theta'} w^i$, and the quotient is
exactly the line bundle $L$.

Let us see what this whole procedure of blowing up amounts to in terms of
the $\sigma$-model gauging. The action (4.1) is obviously invariant under the
transformations (4.3). However, as its zero level set is singular, directly
imposing local invariance and gauging will not be a good attitude. Let us
therefore perform the desingularization first.
 As observed before, this lifts the zero level  an
$\epsilon$ amount. We will consider only the $+\epsilon$ direction since the
other case is less illustrative.

Looking back at the Lagrangian, note that under the change of coordinates
 (4.6), the standard symplectic form on $C^n$, (4.2), becomes
$$-i\sum^n_{i=1} d z^i \wedge d {\bar z}^i =-i [ d w^1 \wedge d {\bar w}^1 +
(\vert w^1\vert^2+\epsilon)\sum^{n}_{i=2} d w^i \wedge d {\bar w}^i
+d(\vert w^1\vert^2) \wedge \alpha ]$$
$$=-i[d w^1 \wedge d {\bar w}^1+\epsilon \pi^* \Omega_{F-S}
+d(\vert w^1\vert^2 \alpha)], \eqno(4.7)$$
where
$$\alpha={1\over 2}\sum^n_{i=2}(w^i d{\bar w}^i-{\bar w}^i dw^i) \eqno(4.8)$$
is the $U(n)$ invariant connection 1-form on $S^{2n-3}$ regarded
as a circle bundle over $CP^{n-2}$, and we have denoted its curvature by
$d\alpha=\pi^* \Omega_{F-S}$, the pull back via the blowing up map
$\pi: L \rightarrow CP^{n-2} $, of the
Fubini-Study 2-form on $CP^{n-2}$. Similarly, the action (4.1) transforms into
$$L=2\int d^2 \sigma \partial_+ w^1\partial_- {\bar w}^1+\sum^{n}_{i=2}
\int d^2 \sigma (\epsilon +\vert w^1\vert^2)\partial_+ w^i
\partial_-{\bar w}^i$$
$$+{1\over 2}\sum^{n}_{i=2}\int d^2 \sigma ({\bar w}^i\partial_+ w^i
\partial_-
\vert w^1\vert^2+\partial_+\vert w^1\vert^2 w^i\partial_- {\bar w}^i),
\eqno(4.9)$$
here we have used the fact that in the new coordinates, the moment map
constraint
(4.5) becomes $\sum^{n}_{i=2}\vert w^i\vert^2=1$. Upon using the equations
of motion,
this can be rewritten as
$$L=2\int d^2 \sigma \partial_+ w^1\partial_-{\bar w}^1+\epsilon
\sum^{n}_{i=2}
\int d^2 \sigma \partial_+ w^i \partial_-{\bar w}^i$$
$$+ ~surface ~~term. \eqno(4.10)$$

The surface term is in complete analogy with the last term of the transformed
symplectic form (4.7). We will assume here that this surface term can be
dropped. However, we should like to mention one interesting situation when this
surface term can not be dropped and plays a prominent role in Floer's study of
symplectic diffeomorphism and holomorphic curves [23].

If we drop  the surface term in (4.10), then the Lagrangian becomes completely
decoupled between $w^1$ and $w^i$, with $w^i$ obeying the constraint
$\sum \vert w^i \vert^2=1$, i.e., constrained on a sphere $S^{2n-3}$. Since
the
$S^1$ action on $w^i$ is free, we can apply the usual steps of gauging the
linear isometry [4], i.e., first promoting to local $U(1)$ transformation
$\theta \rightarrow \theta(x)$, then substituting the minimal coupling of
a gauge field
$$\partial_{\pm}\rightarrow \partial_{\pm}+iA_{\pm}. \eqno(4.11)$$
Solving the equation of motion for the gauge field $A_{\pm}$ in terms of the
1-form constructed from $w^i$:
$$A_{\pm}={i\over 2}\sum({\bar w}^i\partial_{\pm} w^i- w^i\partial_{\pm}
{\bar w}^i) =-{i\over 2}\alpha, \eqno(4.12)$$
and after some trivial manipulations, the final form of the Lagrangian is
$$L=2\int d^2 x\partial_+ w^1\partial_- {\bar w}^1+ \epsilon\int d^2 x \{
{1\over{(1+\vert {\bf w} \vert^2)^2}} {\bf w}\partial_+{\bar{\bf w}}
{\bar{\bf
w}}\partial_- {\bf w} $$
$$- {\delta_{ij}\over{(1+\vert {\bf w}\vert^2)}}\partial_+ w^i\partial_-
 w^j\},~~~~~~i,j=1,..., n-2, \eqno(4.13)$$
where $\vert {\bf w}\vert^2 =\sum \vert w^i\vert^2$, and ${\bf w}
\partial_{\pm}
{\bar{\bf w}}=\sum w^i\partial_{\pm}{\bar w}^i$ (here we have used
inhomogenous coordinates on $CP^{n-2}$). This is a set of $2(n-2)$ real
fields parametrizing $CP^{n-2}$, together with a free, decoupled complex
scalar field.

It is interesting to note the remarkable coincidence of the
gauge field $A_{\pm}$ we introduced
through minimal coupling and the pre-existing 1-form of the Hopf bundle
$S^{2n-3}\rightarrow CP^{n-2}$.

We have chosen to work with a quasi-free $S^1$ action in this section,
while in the last section the reduction carried out could be more general,
i.e. for any (possibly nonabelain) compact group of isometries.
The reason for this
is that, when we treat a general quasi-free $G$ action, it is always
possible to first desingularize the action, passing over to the nearby
regular values of the moment map. Using an observation given in [17] that,
for a regular value of the moment map, $ J: M\rightarrow g^*$,
 the reduced symplectic space with respect to $G$ is equal to the
 corresponding reduced space, with respect to the maximal torus $T$, of
 a submanifold $M'$ of $M$, and for the $T$-action the regular values all
 lie in the interior of some convex polytope, we are able to arrange the
 image set of the moment map so that a one-parameter-subgroup of $T$
 is generated by a non-vanishing element of $t$, which points perpendicularly
 to the wall separating two nearby regions of the polytope of regular
 values. This $S^1$ is what we are discussing in this section.

It is probably worth mentioning that, generalizing the quasi-free group
action of (4.3), to the more general case of (p,q) signature action
(i.e., with p minus and q plus signs in front of $\theta$), the result of
this section gives rise to a sequence of fibred manifolds over $CP^{p-1}$
or $CP^{q-1}$ for negative or positive values of $\epsilon$ parameters,
respectively. Near $\epsilon =0$, the desingularized space looks like
a disc bundle over $CP^{p-1}\times CP^{q-1}$. Passing from $-\epsilon$ to
$+\epsilon$, as explained in [15], thus consists in first blowing up and then
blowing down. The phenomenon is what is called symplectic cobordism.

\vskip 1cm
\noindent {\bf V. The gluing diffeomorphism}

According to the last section, symplectic blowing up can be regarded
locally as gauging
the quasi-free $S^1$ action of a linear $\sigma$-model defined in a small
neighborhood of the singular point, which is diffeomorphic to the linear space
$C^n, n=dim M/2$. To actually accomplish the blowing up of the symplectic
manifold $M$ by gauging, we have to glue the linear blow-up model  back to
the complement of the singular point in $M$ by a certain diffeomorphism
(preserving the symplectic structure). This is accomplished by using a
diffeomorphism on the complement of the singular locus, and the diffeomorphism
arising from the normal coordinate expansion at the singular point.
We will also discuss some subtleties of gluing of the gauge connections.

Let us first of all make a remark  on the role of symplectic diffeomorphisms
in the present context. The usual (bosonic) nonlinear sigma model
with a (Riemannian, or symplectic)
target manifold contains no extra degrees of freedom except those
scalar fields
parametrizing the respective manifolds. The topological sigma
model [23] is one
essential example in which extra dynamical degrees of freedom are introduced
so that it really describes a theory which possesses a high(est) symmetry and
admits, in  a sophisticated way, a generalization to a sigma model with
a symplectic
diffeomorphism. As emphasized by Witten, his theory includes one important
situation where one treats a one parameter family of symplectic manifolds
related by a symplectic diffeomorphism, and which enables him to relate the
global observables of the topological sigma model to the Floer cohomology
group. Now, in our case
a one parameter family of symplectic manifolds certainly arises in the
process of blowing up as in Section 4. These are Marsden-Weinstein reduced
spaces with respect to different values of the
 moment map other than zero. However, as we have mentioned in Section 2, the
linear sigma model fields cannot be treated as independent degrees of
freedom of the whole symplectic sigma model, rather, it is
related to the nonlinear sigma model by a certain diffeomorphism. Thus, the
situation in which we are involved here is actually a symplectic nonlinear
sigma model
together with a diffeomorphism. It is quite plausible that by going over to
the symplectic generating space (or a cobordism) for a family of reduced
spaces, as described in [15], one can find a diffeomorphism invariant theory
which effectively calculates the global transition between different geometric
and topological structures on both sides of the blown up. Leaving aside this
interesting possibility, we restrict ourselves in the present paper only to
the case
where we treat the symplectic diffeomorphism as gluing data for forming the
globally defined blown up sigma model. We will first take a closer look at the
diffeomorphism on the complement of the singular point, which is used to glue
together the linear symplectic forms, and then use the exponential
mapping to discuss some further properties of the symplectic diffeomorphisms.

\noindent {\it Diffeomorphisms on $L-$(zero  section)}

As we have seen in the last section,
the blowing up of the origin in the linear space $C^n$ results in a symplectic
form
$\epsilon \pi^*\Omega_{F-S}$ defined in a small neighborhood of the blown up
point,
where $\pi$ is (one part of) the blowing up map, $\pi: L \rightarrow
CP^{n-1}$,
the other part of this map is called $\beta: L \rightarrow C^n$, which maps
the zero section into one point (the origin) in $C^n$. On the complement of
the
zero section, i.e., on $L-$(zero section), the map $\beta$ is a bijection,
or in
general terms, a G equivariant symplectic diffeomorphism. The role of this
diffeomorphism is to pull back the G-invariant symplectic form on $C^n-(0)$
to $L$ and glue it to the blown up $\pi^* \Omega$. Let us see how to construct
symplectic diffeomorphisms in the linear case.

Remember we have obtained our
blown up manifold (a line bundle over $CP^{n-1}$) as a Marsden-Weinstein
reduced space with respect to a nonzero value $\epsilon$ of the moment map.
The value $\epsilon$ lies in the dual of the Lie algebra of $S^1$, i.e., $R$.
Now the moment map $\mu$ of the $S^1$ action does not uniquely correspond to
 the symplectic 2-form
on $C^n-(0)$, but leaves undetermined an arbitrary constant. Indeed [15], let
$f(s),  s=\vert z\vert^2$ be an arbitrary diffeomorphism of $R^+\subset R$.
Viewed as a function of $z$, $g(z)=f(\vert z \vert^2)$, an arbitrary $U(n)$
 invariant symplectic form on $C^n-(0)$ has the form
$$-i\partial{\bar \partial} g=-i[f''(\vert z\vert^2){\bar z}dz\wedge z
d{\bar z}
+f'(\vert z\vert^2)dz\wedge d{\bar z}]. \eqno(5.1)$$
By taking the interior product of this form with an $S^1$-valued vector field,
we obtain
the moment map in terms of the function $f$,
$$i(\xi)[-i\partial{\bar \partial} g]=d\mu(\vert z\vert^2),$$
where
$$ \mu'(s)=f'(s)+sf''(s), \eqno(5.2)$$
or
$$\mu(s)=sf'(s)+c_1. \eqno(5.2a)$$
The solution of this equation for $f$ in terms of $\mu$ contains therefore as
integration constants
$$c_1+c_2 \log s.$$
This means that any two closed 2-forms with the same moment map differ by a
 constant multiple of the form
$$\Omega_{F-S}=-i\partial {\bar \partial} \log (\vert z\vert^2). \eqno(5.3)$$
In particular, the linear symplectic form (corresponding to the choice
 of $f=s$), when perturbed by a term $\epsilon \Omega_{F-S}$, is a well
defined,  $U(n)$ invariant
symplectic form on $C^n-(0)$. The question then arises what symplectic
diffeomorphism
can bring this symplectic form into the corresponding one on $L$. It
turns out [15]
that there always exists such a symplectic diffeomorphism if there is
a diffeomorphism
$h$ of $R^+$ onto itself such that the integral constant $c_1$ in (5.2a)
 is left
invariant. Then the required symplectic diffeomorphism is
$$\psi(z)=h(\vert z\vert^2)z. \eqno(5.4)$$

A simple calculation reveals that $h(\vert z\vert^2)$ can be taken to be
$$h(s)=(1+\epsilon (\sigma -1)s^{-1}+\epsilon \sigma'\log s)^{1\over 2},
\eqno(5.5)$$
then,
$$\psi^* \omega_1 =\omega_2,$$
$$\omega_1 =-i \partial{\bar \partial} (s+\epsilon \log s),$$
$$\omega_2= -i \partial{\bar \partial}(s+\epsilon \sigma (s) \log s ),
\eqno(5.6)$$
where $\sigma (s)$ is a non-negative smooth function of compact support
(a cut-off function) which is identically one in some neighborhood of
the origin.

This consideration can be generalized to the case where the quadratic moment
map $\mu(z)$ is replaced by $\phi(z)=f(z)/\vert f(z)\vert$, with $f(z)$ a
polynomial function of n-complex variables $z=(z_1, z_2,..., z_n)$. $\phi(z)$
defines a mapping from the zero level set of $f$ to $S^1$. According to a
theorem of Milnor [25, 26], in the complement of a small neighborhood of the
singular point in the zero level set, $\phi(z)$ is a projection map of a
 smooth fiber bundle over $S^1$.

\noindent{\it Normal Coordinate Expansion}

As  observed in the last subsection, a symplectic diffeomorphism associated
to the blowing up map usually arises as a consequence of the indeterminacy
of the G-equivariant
closed 2-form by the moment map. Globally this indeterminacy gets fixed
by specifying  the symplectic invariants, such as the additive constant $c_1$
and the interval of the image of the moment map. Then the globally
defined symplectic diffeomorphism
is given by some diffeomorphism of $R^+$ onto a subinterval $I$.

The aim of this subsection is to use the method of normal coordinate
 expansion [27], [19] to clarify some further properties of the symplectic
diffeomorphism in the context of blowing up. Application of this method
 to multi-loop calculations
in the nonlinear sigma model has been initiated by Honerkamp
 in the early 70's
[28], and considerably elaborated by Alvarez-Gaum\'e et al a decade later [29].
It is useful to note that whereas the references [28, 29] deal with
 perturbation expansions of the quantum theory, therefore specifying to a
 situation where the
normal coordinate variables are treated as quantum fields, as opposed to the
fields sitting over the origin of the normal coordinate system which are
treated
classically, we will here treat the normal coordinate variables simply as
a change of coordinates in a small neighborhood containing the singular point.
In this way we are able to provide a local version of our blowing up
diffeomorphism.
Together with the global one, given by (5.4) in the last subsection,
this completes the whole set of (symplectic) diffeomorphisms needed to glue
the blow-up back to the global reduced space.

In the following, we discard the topological term from our discussions and
consider the term containing the metric tensor only.
This is reasonable because in the normal
coordinate expansion, everything is expressed in terms of solutions of the
 geodesic
equation, and it does not matter whether we use the Riemannian connection or
a general affine connection with torsion.
We choose a small neighborhood ${\cal U}$ of
an arbitrary point $p$ in $M$, such that any two points in ${\cal U}$ are
connected by a single geodesic. Thus, there exists a coordinate
transformation from the coordinate system $x^i$ with origin $x^i_0$ being
the coordinates of the point $p$, to $u^i$ with $u^i_0
=0$ as the origin in the tangent space at $p$.
This transformation is given by the solution of the geodesic equation
$${{d^2 x^i}\over{d t^2}}+\Gamma^i_{jk} {{d x^j}\over{dt}}{{dx^k}\over{dt}}
=0, \eqno(5.7)$$
here $\Gamma^i_{jk}$ is the affine connection on $M$, (note that only the
symmetric part of this connection appears in this equation,)
and $x^i(t)\vert_{t=0}=x^i_0,~~ dx^i(t)/dt \vert_{t=0}=u^i$, as initial
conditions. The solution $x^i(t)$ in terms of the initial values can be
 expressed as a Taylor series near $x^i_0$
$$x^i(t)=x^i_0+u^i t- {1\over 2}\Gamma^i_{jk}u^ju^k t^2- .... \eqno(5.8)$$
with higher power terms containing the derivatives of $\Gamma^i_{jk}$.

Since the Jacobian $(\partial x^i/\partial u^i)_{u_0=0}=\delta^i_j$, the
transformation is regular at the origin $u^i_0=0$. The coordinate system
$(u^i)$ in a small neighborhood of $p$ is called the normal coordinate system.
The map between the two coordinate systems is the exponential map
we have mentioned
before. The exponential mapping is in fact a local diffeomorphism  in
the sense that
there exists a neighborhood ${\cal U}$ of a point $x_0$, which is contained
in the normal coordinate neighborhoods of all points in ${\cal U}$.
With $x^i$ and $u^i$
viewed as the scalar fields in the $\sigma$-model, the actions in terms
of the field $u^i$ and of $x^i=x^i(t=1)$ possess the same form up to a
redefinition of the metric tensor. Thus when we do not think of $x^i_0$ as a
classical field upon which  one was to perform the background field expansion,
passing from $x^i$ to $u^i$ simply amounts to a local change of coordinates,
 with  $x^i_0$ being constant, thus disappearing from the action.
Now consider the situation where we have a nonlinear sigma model with
a (symplectic)
diffeomorphism, then generally the statement that $x^i_0$ is a constant
configuration remains true only if $x^i(t)\vert_{t=0}=x^i_0$ is preserved by
the diffeomorphism $\psi$, i.e., if the  point $p$ is a fixed point of $\psi$.
When $p$ is not a fixed point of $\psi$, $x^i_0$ acquires a nontrivial
dependence
on spacetime via the diffeomorphism $\psi$. In this case, if we perform the
normal coordinate expansion for the action, we get [29]
($\phi=\psi^*(x^i_0)$),

$$S=\int d^2 \sigma G_{ij}\partial_{\mu}x^i\partial^{\mu}x^j=S_0+S_1$$
$$=\int d^2 \sigma G_{ij}\partial_{\mu} \phi^i\partial^{\mu}\phi^j
+\int d^2 \sigma G_{ij}\partial_{\mu}\phi^i D^{\mu}u^j$$
$$+{1\over 2}\int d^2 \sigma
(G_{ij}D_{\mu}u^iD^{\mu}u^j
+R_{ijkl}u^k u^l \partial_{\mu} \phi^i \partial^{\mu} \phi^j)$$
$$+ {\cal O}(u^3), \eqno(5.9)$$
where
$$D_{\mu}u^i=\partial_{\mu} u^i+\Gamma^i_{jk} u^j\partial_{\mu}\phi^k.$$
Note that we have used the fact that under a generic coordinate change, the
$u^i$ transforms as  a contravariant vector while the form of the geodesic
equation (5.7) remains unchanged if the diffeomorphism $\psi$ is such that
it reduces to a linear function of $t$ when restricted to the geodesic curve
$t\rightarrow x(t)$.
The above expansion is valid at any point of the manifold $M$. Especially,
it is valid at the singular point when it is not a fixed point of the
diffeomorphism $\psi$.

Now we can see the role of the normal coordinate expansion. It embodies the
diffeomorphism $\psi$ on the large (which reduces to a linear function of
the geodesic length) as new dynamical degrees of freedom which
disappear when the singular point is a fixed point of $\psi$, and provides
a (weakly) coupled form of the total action (5.9). It is weakly coupled in
the following sense: the coupling between $u^i$ and $\phi^i$ in (5.9) is in
a form such that it contributes to the one-loop divergences of the theory and
must be  decoupled when we add the one loop renormalization
counter terms of the form
$$ {1\over {4\pi \epsilon}}\int d^2 \sigma [R_{ij}\partial_{\mu}\phi^i
\partial_{\mu}\phi^j +G^{jk}\Gamma^i_{jk}{{\delta S_0}\over{\delta \phi^i}}].
\eqno(5.10)$$
Then the action is completely decoupled modulo equations of motion by the
following equality (which is easily proved):
$$-{{\delta S_0}\over{\delta \phi^i}}u^i=\int d^2 \sigma G_{ij}
\partial_{\mu}\phi^i D^{\mu}u^j. \eqno(5.11)$$

Thus, the remaining total action, when the field $u^i$ is lifted to the
 tangent
frame by using the $m$-bein, $e^a_i,~~a=1, . . . , m$, can be put into
the form
$$\int d^2\sigma G_{ij}\partial_{\mu}\phi^i \partial^{\mu}\phi^j
+{1\over 2}\int d^2 \sigma D_{\mu}u^a D^{\mu} u^a. \eqno(5.12)$$

Note that the action for $u^a=e^a_i u^i$ is already gauge invariant
 with a maximal gauge invariance $SO(m)$ and the gauge field
$(\omega_i)^{ab}
\partial_{\mu}\phi^i=A^{ab}_{\mu}$, is matrix valued in $so(m)$.
It is the pull
back to spacetime of a Yang-Mills connection on the (orthogonal) frame
bundle over $M$, i.e., a one-form $A^{ab}_i d\phi^i$.

\noindent{\it Reduction of Connections}

The connection form appearing in (5.12) for the linear model is apparently
$so(m)$ valued, but usually we can not gauge the maximal isometry group,
instead, for the linear model in the last section,
only  the $U(1)$ subgroup is to be gauged. To carry out gauging by
associating a gauge field to the sigma model is thus a problem of reducing
the gauge connection of an orthogonal frame bundle to that of a subbundle of
 the bundle of
unitary frames. Here we discuss in what sense the connection which appeared
above is related to the gauge field arising from gauging.

It is well known that if the Hamiltonian group action on the symplectic
manifold is free, then the regular level set is a principal fiber bundle
over the reduced symplectic space. Any connection can be used to define the
integrable distribution which  is equivalent to the tangent space
 decomposition.
 In the case of a singular reduction, instead of a fiber bundle,
one has a stratified space, each of whose strata is a regular
 Marsden-Weinstein
reduced space, under a rather general assumption [30].
In the case of the point singularity, there are only two strata:
the complement
of the singular point $p$ in $M$, and $p$ itself. The isotropy group at
$p$ is the same as the subgroup of the isometry group we choose to
gauge. It follows from the results of [30] that the fiber types of both the
global regular reduced space and the blown up at the singular point
$p$ are the same, hence their respective principal $G$-bundles admit the same
decomposition of tangent vectors. Thus the following discussions for the
linear
case are equally applicable to the nonlinear case with minor modifications.

Given a connection on a principal fiber bundle $P(M,G)$, which is the same
as given a direct sum decomposition of the tangent space at $x\in P$
$$T_xP=Vert \oplus Hor, \eqno(5.13)$$
into the  vertical and horizontal subspaces (for details see the second
reference of [19]). A connection $\Gamma$ on $P(M,G)$ is  reducible to
a connection $\Gamma'$ on a subbundle
$Q(M,H)$ with $H\subset G$, if for  any point $x$ on $Q$, the horizontal
subspace $Hor(P)$ of $T_x P$ is tangent to $Q$. And if $\phi$ is a
homomorphism of groups
$\phi: G \rightarrow H$, (which induces a homomorphism on the Lie algebra
level, still denoted by $\phi$,)
then $\phi\cdot \omega=\phi^* \omega'$, where
$\omega, \omega'$ are connection 1-forms with respect to $\Gamma, \Gamma'$.

In the cotangent space approach for connections, in terms of Lie algebra
valued 1-forms, a connection 1-form admits the general expression:
$$\omega=\omega_0+g^{-1}Ag=g^{-1}dg+g^{-1}A_{\mu}gdx^{\mu}, ~~g\in G
\eqno(5.14)$$
where $A=A^a_{\mu}\lambda_adx^{\mu}$, and $\omega_0$ is the Maurer-Cartan
canonical 1-form. At the point $(1, x)$ on the locally trivialized
principal $G$-bundle $P_{\alpha}=G\times{\cal U}_{\alpha}$, the tangent
space $T_{(1,x)}P_{\alpha}$ is spanned by vectors $(A_{\mu}, \partial/
\partial x^{\mu}=\partial_{\mu})$, with $A_{\mu}$ viewed as an element of
the Lie algebra of $G$. An arbitrary tangent vector in $T_{(1,x)}
P_{\alpha}$, $X=X_H+X_V$ can be expanded in terms of the basis
$(A_{\mu}, \partial_{\mu})$. Now we can deduce the reduced connection on the
subbundle $Q(M,H), ~H\subset G$ as follows. Rewrite $\phi\cdot \omega^G=
\phi^*\omega^H$ as $\phi(\omega^G(X_V))=\omega^H(\phi(X_V))$. Let $g=h\oplus
m$ be the direct sum decomposition of the Lie algebra of $G$, therefore
$\phi: g\rightarrow h$ is a projection. We have two conditions determining
$\omega^H$ uniquely from $\omega^G$, the first implies that, $\omega^G(X_V)
\vert_h =\omega^H(X'_V)$, where $X'_V$ on $T_{(1, \phi(x))}Q$ are fundamental
vector fields corresponding to elements of the Lie algebra of $H$. The second
says simply that certain vertical components $X_V$ are mapped by $\phi$
into some linear combinations of horizontal vectors of $TQ$. Using the
tangent basis $(A_{\mu}, \partial_{\mu})$, the first condition gives
$$\omega^G(A_{\mu},\partial_{\mu})\vert_h =A_{\mu}\vert_h =A'_{\mu}
=\omega^H(X'_V)=\omega'(X'_V),$$
and the second condition just expresses the fact that $(A'_{\mu},
\partial_{\mu})$ can be chosen as the basis for the horizontal subspace
of $TQ$. Thus the horizontal part of the connection 1-form on the
reduced principal bundle is (using the projection $S:T_yQ \rightarrow T_yM$)
$$S\omega'(A'_{\mu},\partial_{\mu})=\omega'(S(A'_{\mu},\partial_{\mu}))
=g^{-1}A'_{\mu}g, ~~~g\in H. ~\eqno(5.15)$$
Note that its curvature is by definition horizontal: $\Omega=Sd\omega=
g^{-1}(dA+[A,A])g$.

The reduction process concerned here corresponds to first passing from
the $SO(2n)$ bundle of the linear frames to the $U(n)$ bundle of the complex
linear frames, and then reducing the $U(n)$ bundle to the abelian $U(1)$
(or rather the maximal torus $T$) sub-bundle. As the
symplectic manifold $M$ has an almost complex structure, its tangent space
is equipped with the standard complex structure $J$, which enables us to
complexify the coordinates on $T_xM \sim C^n(z=x+iy, {\bar z}=x-iy)$. The
complexification induces naturally a direct sum decomposition of the Lie
algebra $so(2n)=u(n)\oplus m$ with $m$ the orthogonal complement of $u(n)$
in $so(2n)$. According to our discussion above,
we see that the (horizontal part of the) connection 1-form on the
$U(n)$-bundle as reduced from that of the $SO(2n)$ bundle is simply
$$A^{a}=A^a_idz^i+A^a_{\bar{i}}d\bar{z^i},$$
where $a$ is the index for the basis of the Lie algebra of $U(n)$, and
$i,\bar{i}$ run from $1$ to $n$.

For the reduction of  the $U(n)$ connection to the $H=U(1)$ connection, it
suffices to consider the reduction to $T=U(1)^n$, the maximal torus of
$U(n)$, since from $T$ to $U(1)$ the process is simply by taking diagonals.
Thus assume the Lie algebra $u(n)$ admits the decomposition $u(n)=t\oplus
m$, let $\phi: u(n)\rightarrow t$ be the projection. $\phi$ maps the
horizontal tangent vectors to horizontal vectors. Denote an arbitrary
tangent vector in $u(n)\times T_{(z,{\bar z})}{\cal U}_{\alpha}$ as
$\tau=(A_i+A_{\bar i}, \partial_i+ \bar{\partial_i})$, therefore
$$\phi: \tau=(A_i+A_{\bar i}, \partial_i+ \bar{\partial_i}) \rightarrow
(\phi(A_i)+\phi(A_{\bar i}), \partial_i+\bar{\partial_i}) ~ \in t\times
T_{(z, {\bar z})} {\cal U}_{\alpha}.$$
We thus have the following equations determining the unique $T$-connection:
$$\omega^T(~~\cdot~~, \partial_i+ \bar{\partial_i})=\phi(A_i)+\phi(A_{\bar i}
)=A'_i+A'_{\bar i}$$
$$\omega^T(\xi^i A'_i+\bar{\xi^i}A'_{\bar i}, ~~\cdot~~)=0,
{}~~A' \in t,  \eqno(5.16)$$
where $~\cdot~$ means "for any" vertical (horizontal) components of the
tangent vector, $\xi^i, \bar{\xi^i}$ are arbitrary holomorphic
(anti-holomorphic) functions of $(g, z)$. From the last equation of (5.16),
we deduce that
$$g^{-1}A'_i gdz^i+g^{-1}A'_{\bar i}gd\bar{z^i}=\omega_0=g^{-1}dg,
{}~~g \in T. \eqno(5.17)$$
If we choose the parametrization of the group $T$ as follows
$$T=U(1)^n=\{(z^1,z^2,...z^n)\in (C^*)^n \vert~~ \vert z^i\vert^2=1\},$$
it is not difficult to see that the connection thus obtained coincides, for
the diagonal $U(1)$, with the one which has appeared in (4.12).
This completes our discussion of reduction of the connection in the linear
$\sigma$-model. On the other hand, the general connection used to gauge the
nonlinear $\sigma$-model belongs to the subgroup of the whole isometry
group of its tangent space. By the same arguments as above, one can deduce
that it can be reduced to the connection of the principal $T$-bundle,
with $T$ the maximal torus of the gauge group $H$. This, when expressed
in terms of the $\sigma$-model scalar fields, is related to the connection
of the linear $\sigma$-model precisely by the gluing diffeomorphisms
discussed before.

We now comment briefly on the process of integrating out gauge fields. It is
known that this process receives quantum corrections at the sigma model
loop level. Since the one loop effect has been vital in our discussion of
the normal coordinate
expansion, it is expected that it is also important to include the
quantum corrections in the blowing up construction. Unlike the case of the
conformally invariant sigma models where the one loop corrections can be
conveniently summarized into the dilaton shift, in the present situation,
there is no place for a dilaton, neither the particular notion of conformal
 invariance,
but instead, we have the symplectic diffeomorphism which might have
nontrivial
fixed point structures. It may happen that these fixed point structures
 manifest themselves
into some unforseeable dynamical modes of the (gauged) nonlinear sigma model,
e.g. the appearance of the kinetic term for the gauge fields, in much
like the way it arises  in ref[31], where an interesting
mechanism for generating dynamics for the gauge field by the $1/N$ corrections
to the $CP^N$ model is suggested. We leave the discussion of
this possibility for future work.
\vskip 1cm
\noindent{\bf VI. Applications}

\noindent{\it Toric $\sigma$-models}

A toric manifold associated with an integral or rational polyhedron can be
obtained as a symplectic reduction of $C^N$ by the Hamiltonian action of
the subtorus of $T^N_C=(C^*)^N$, at a regular level of the corresponding
moment map (see [16] for more properties of the toric manifolds). Obviously
a toric $\sigma$-model (i.e. a $\sigma$-model with target space being a
toric manifold) can be viewed as a suitable symplectic reduction of the
linear $\sigma$-model, or as the gauged nonlinear $\sigma$-model studied
in section III. It is a nontrivial fact that some Hamiltonian subtorus
actions on the toric manifold can have fixed points when the image of these
points are exactly the vertices of the convex polyhedron. We are interested
in the effect of blowing up a point in the toric $\sigma$-model.

Let us take the simplest example of $CP^2$, constructed as a toric manifold
whose associated polyhedron is the standard 2-simplex, i.e. a triangle
$\Delta\subset R^2$.
If $e_1, e_2$ denote the basis vectors of $(Z^2)^*
\subset R^2$, which are two edge vectors of $\Delta$, we can form a fan whose
1-skeletons (edges of the 2-cones) are all of the form
$tx_i,~0\le t<\infty, ~i=1,2,3$, where
 $x_i=e_i,~ x_3=-(e_1+e_2)$. One can take $x_i$ to be the basis vectors
in $Z^3$, thus there exists a natural map $Z^3\rightarrow Z^2, ~ {\bf x}
\mapsto {\bf e}$ which induces the corresponding map $R^3\rightarrow
R^2$ and the quotient map $T^3\rightarrow T^2\rightarrow 0$ with kernel
$S^1$. The realization of $CP^2$ as symplectic reduction of $C^3$ is carried
out by reducing $C^3$ by the (smooth) Hamiltonian action of this $S^1$.
{}From this construction it is obvious that $T^2\subset T^3 $ acts on
$CP^2$ in a Hamiltonian fashion and the image of the $CP^2$ under its
moment map is exactly $\Delta$. (The same steps work for other simple
toric mainfolds, giving rise to Hamiltonian spaces of dimension twice of
that of the corresponding torus.)

We already know that the nonlinear $\sigma$-model of $CP^2$ can be expressed
as a gauged $\sigma$-model with the gauge field
$$A_{\mu}={i\over 2}\sum_{i=1}^3 {\bar z}^i \partial_{\mu} z^i
-z^i\partial_{\mu}{\bar z}^i, \eqno(6.1)$$
and the action of the form $\sum D_{\mu}z^i \overline{D_{\mu} z^i},~ D_{\mu}
=\partial_{\mu}+iA_{\mu}$. The integral of the symplectic form over a
homology cycle in $CP^2$ equals
 a topological invariant $ 1/2\pi \int \epsilon_{\mu\nu}
\partial_{\mu}A_{\nu}$ which is the first Chern number of the tangent bundle.
The $CP^2$ $\sigma$-model has a global $SU(3)$ invariance, of which the
maximal torus $T^2$ acts in the Hamiltonian fashion. We know that this $T^2$
action is not free at some points whose image under the moment map are the
vertices of $\Delta$. This may cause serious problems in the quantum theory,
eventhough it is harmless classically, as far as one does not perform the
quotient (which is a point in this case).
A possible resolution is to blow up the fixed point on $CP^2$.
Application of our general procedure leads to a nonlinear $\sigma$-model
whose target space may be identified as a connected sum of $CP^2$ and
another $CP^2$, considered as the projectivization of a line bundle over
$CP^1$, with the symplectic form of the latter $CP^2$ multiplied by
a small real number $\epsilon$. One implication of this example is, that
quantum mechanically, the blown up $CP^2$ model automatically overcomes
the zero area limit, as the transition from $-\epsilon$ to $+\epsilon$
in this construction is smooth. Details of the blown up toric $\sigma$
model will be reported separately.

\noindent{\it N=2 Supersymmetric $\sigma$-model}

In this case, there exists a general procedure [3,6] of performing the
$N=2$ quotient by gauging the (holomorphic) isometries of the $N=2$
superspace action of the form
$$S={1\over 2}\int d^2\sigma D_+ D_-{\bar D}_+{\bar D}_- K(\Phi,
{\bar \Phi}, \Lambda, {\bar \Lambda}), \eqno(6.2)$$
with arbitrary chiral and twisted chiral superfield multiplets $\Phi,
\Lambda$. The gauged action takes the general form of a new K\"ahler
potential $K'$ which is the original potential $K$ with $\Phi$ and
$\Lambda$ minimally coupled to some gauge multiplet $V$, plus terms
 which are trivially gauge invariant, such as the Fayet-Iliopoulos terms
 which are present when the isometry group contains $U(1)$ subgroups.
 In the same spirit as the bosonic $\sigma$-model and its symplectic
reduction studied in the previus sections, we can carry out the symplectic
blowing up for the $N=2$ $\sigma$-model as well. The basic ingredients
are a superspace analogue of the normal coordinate expansion on the one hand,
and the identification (and the interpretation) of the blowing up parameter
$\epsilon$ in the linear model as the coupling constant in front of the
Fayet-Iliopoulos terms, on the other hand. We will not describe both these
important points here. However, the picture of the blown up $N=2$
$\sigma$-model is clear: to each $N=2$ supersymmetric $\sigma$-model,
arising from gauging an appropriate holomorphic isometry group, at any
(isolated) configuration which is the fixed point  of a subgroup of the
isometry group,
one can associate a gauged version of the $N=2$ linear $\sigma$-model.
The linear $N=2$ $\sigma$-model has been studied intensively in [22].
Ref [32] contains also discussions of the flops in the Calabi-Yau spaces
which are in fact a sequence of blow-ups and blow-downs.

\vskip 1cm
\noindent{\bf VII. Summary and conclusions}

What has been done in the previous sections can be actually viewed as
giving concrete explanations to the various pieces of information contained
in the following formula for the symplectic form on the reduced space
which is diffeomorphic to a connected sum of the global reduced space and
a copy of $CP^n$:
$$\Omega_{\epsilon}=\Omega_0+\epsilon\pi^*\Omega_{F-S}. \eqno(7.1)$$
Perhaps we should add that it is a well-defined quantity.
Although tedious, it can be checked directly that the
2-form $\Omega_0$ as obtained from (3.27), after integrating out gauge
fields, is closed. From the arguments given before (5.9) it follows that
the fields in the nonlinear $\sigma$-model have no support near the singular
locus. The connected sum is formed by gluing the linear $\sigma$-model back
to the gauged nonlinear one along an annular region (smoothly) diffeomorphic
to $B_{2\epsilon} -B_{\epsilon}$  of balls in $C^n$. Thus the reduced
symplectic form is well-defined on the blow-up.

{}From the symplectic geometric point of view, blowing up is a useful tool
to obtain numerous interesting examples of symplectic manifolds [33].
The symplectic forms are classified into the  equivalence classes
up to diffeomorphisms of certain type. It is a challenging problem to
calculate some invariants on the space of equivalence classes of symplectic
forms physically. For example, a generalization of the
Duistermaat-Heckman theorem to the singular reductions [16] states that
the cohomology class of $\Omega_{\epsilon}$, $[\Omega_{\epsilon}]$ is
continuous as an affine function of the parameter $\epsilon$ (even at
$\epsilon=0$), the slope of the line segment in $H^2(M)$ spreaded by
$[\Omega_{\epsilon}]$ for all $\epsilon < 0$, gets a jump when going
through $\epsilon=0$. It is hoped that the results of this paper may
point a way to effectively calculating this interesting invariant.

We have suggested in this paper a gauged nonlinear sigma model for the
symplectic blowing up, and discussed its various aspects as a well defined
model for describing the non-singular symplectic manifold resulting from
blowing up a singular point. The model consists of gauging  both nonlinear
and linear parts of the action, resulting in the respective symplectic
quotients. The linear sigma model has been used to provide a local version
of the symplectic blowing up (which is in the linear case identical
to the corresponding complex blowing up, i.e. passing from the affine
varieties to the projective ones), whereas the nonlinear part of the action
describes the complementary region of the singular point of the  target
manifold. The two pieces of the  construction are glued together by the
symplectic diffeomorphism which appears as part of the blowing
up maps (the birational morphisms). Our main conclusions are, the
symplectic blowing up is a process that is well defined in a  gauged
nonlinear sigma model with symplectic diffeomorphism, and the blow-up
which is diffeomorphic to a connected sum of a symplectic manifold and
a copy of the complex projective space, is described by the resulting
classical configurations (instantons).

One motivation of the present work is to try to understand in physical terms
what is involved in the construction of the symplectic cobordism described in
[15]. While it is relatively easy to convince oneself that the gauged sigma
model is the appropriate arena here, it is far from trivial to identify the
blow up modes in the gauged sigma model. Classically, it turns out, the
blow up modes can be viewed as the consequence of the existence of the
symplectic diffeomorphism. Quantum mechanically, it is quite plausible
that the fluctuation around the classical blow up modes might be much
enhanced, or in the language of the sigma model, the instanton corrections
might become more significant. The advantage of working with the
nonlinear sigma models is that here the instanton effects can be
conveniently handled, as they have long been explored. Another aspect of
the result is that it seems to provide a concrete construction of the
so-called topology changing process in terms of the nonlinear sigma model.
Comparing with the recent work [7, 32], our result seems to bring together
the local and global analyses separately pursued by those authors.
\newpage
\noindent {\bf Acknowledgements}

H.B.G. thanks the Alexander von Humboldt Foundation for financial
support.    We thank Dr. Thomas Filk for a careful reading
 of the manuscript.
\vskip 1.2cm
\noindent {\bf Appendix }

We present in this Appendix the proof that there exists a change of
coordinates so that
in a small neighborhood of a point in $M$, any smooth vector takes the form
(3.18) or (3.19).

Let an arbitrary smooth vector in an arbitrary coordinate system in which
it is nonvanishing be
$$X=\sum^m_{i=1}\xi^i{\partial\over{\partial u^i}}. \eqno(A1)$$
Because it is nonvanishing, at least one of its components cannot be zero.
Let $\xi^1$ be a nonzero component. Consider the differential equations
$${d u^{\alpha}\over{d u^1}}={{\xi^{\alpha}(u^1,..., u^m)}
\over{\xi^1(u^1,..., u^m)}}
,~~~ 2\leq \alpha \leq m, \eqno(A2)$$
with $u^{\alpha}$ arbitrary functions (of variable $u^1$). According to
the theory
of ordinary differential equations [34], there exists a unique set of
solutions of (A2) in a sufficiently small neighborhood, $\vert u^1 \vert
< \delta$ in $\cal U$
$$u^{\alpha}=\phi^{\alpha}(u^1), ~~~~ \vert u^1 \vert < \delta,
\eqno(A3)$$
obeying the  prescribed initial conditions
$\phi^{\alpha}(0)=v^{\alpha}$. $\phi^{\alpha}$ depend smoothly on
 $u^1$ and the initial values
$v^{\alpha}$, therefore can be taken as functions $\phi^{\alpha}
(u^1, v^2,... v^m)$.
Make the change of coordinates
$$u^1=v^1$$
$$u^{\alpha}=\phi^{\alpha}(v^1, v^2,..., v^m), ~~~ 2\le \alpha \le m.
 \eqno(A4)$$
Because the Jacobi is equal to 1 at $v^1=0$, there is a coordinate
 neighborhood
${\cal V}\subset {\cal U}$, $\{{\cal V}; v^i\}$, such that
$$X \vert_{\cal V}=\sum \xi^i{\partial\over{\partial u^i}}=
\xi^1{\partial\over{\partial u^1}}+\sum^m_{\alpha=2}\xi^{\alpha}{\partial\over
{\partial u^{\alpha}}}$$
$$=\xi^1\sum^m_{i=1}{{\partial u^i}\over{\partial v^1}}{\partial\over{\partial
u^i}}=\xi^1{\partial\over{\partial v^1}}. \eqno(A5)$$
Thus by defining
$$w^1=\int^{v^1}_0 {{dv^1}
\over\xi^1}, ~~~ w^{\alpha}=v^{\alpha}, ~~~ \alpha=2,. . . , m, \eqno(A6)$$
in the new coordinate system $({\cal W}; w^i)$,
the vector $X$ takes the form
$$X\vert_{\cal W}={\partial\over{\partial w^1}}. \eqno(A7)$$

We can perform a chain of changes of coordinates, until all the Hamiltonian
            vectors $\xi_a$ are transformed into the form (A7). To write
 out the general
formulae, note that in the case of two vectors $\xi_{(1)}, \xi_{(2)}$,
in the coordinate
system in which $\xi_{(1)}$ has the form $\xi_{(1)}=\partial/\partial
w^1_{(1)}$,
due to the linear independence of $\xi_{(1)}$ and $\xi_{(2)}$, the
 coefficients
of $\xi_{(2)}$ can not depend on $w^1_{(1)}$, thus
$$\xi_{(2)}=\sum^m_{\alpha=2}\xi^{\alpha}_{(2)}{\partial\over{\partial
w^{\alpha}_{(1)}}}= .~.~.~      $$
$$=\xi^2_{(2)}{\partial\over{
\partial w^2_{(2)}}}, \eqno(A8)$$
when
$$\xi^{\alpha}_{(2)}/\xi^2_{(2)}=d w^{\alpha}_{(1)}/d w^2_{(1)}
=d w^{\alpha}_{(1)}/d w^2_{(2)} \eqno(A9)$$
are satisfied by the new coordinates $w^i_{(2)}$.

In general, after a chain of coordinate changes
$$w^i_{(0)}\rightarrow w^i_{(1)}\rightarrow .~.~.~
\rightarrow w^i_{(a)},$$
the vectors $\xi_a$ are transformed into the form
$\{ \partial/\partial w^1_{(a)}, \partial/\partial w^2_{(a)},. . . ,
\partial/\partial w^a_{(a)}\}$. It is easy to derive the following relations
$${{\partial w^s_{(s-1)}}\over{\partial w^s_{(s)}}}=\xi^s_{(s)}, ~~~
{{\partial w^{\alpha}_{(s-1)}}\over{\partial w^s_{(s)}}}=\xi^{\alpha}_{(s)},
  ~~~
{{\partial w^{\alpha}_{(s-1)}}\over{\partial w^{\beta}_{(s)}}}=
\delta^{\alpha}_{\beta}, $$
$$ s=1, .~.~.~, a; ~~~~~ \alpha, \beta=s+1, .~.~.~, m. \eqno(A10)$$
\newpage
\noindent {\bf References}
\begin{enumerate}
\item L. Alvarez-Gaum\'e and D.Z. Freedman, Commun. Math. Phys. {\bf 80},
443(1981).
\item J. Bagger and E. Witten, Phys. Lett., {\bf 118B}, 103(1982);\\
J. Bagger, Nucl. Phys. {\bf B211}, 302(1983).
\item C.M. Hull, A. Karlhede, U. Lindstr\"om and M. Rocek, Nucl. Phys.
{\bf B266}, 1(1986).
\item N.J. Hitchin, A. Karlhede, U. Lindstr\"om  and M. Rocek, Commun.
Math. Phys. {\bf 108}, 535(1987).
\item E. Witten, Phys. Rev. {\bf D44}, 314(1991);
\item M. Rocek and E. Verlinde, Nucl. Phys. {\bf B373}, 630(1992).
\item P. Aspinwall, B. Greene and D. Morrison, Phys. Lett. {\bf 303B},
249(1993);\\
 E. Witten, Nucl. Phys. {\bf B403}, 185(1993);
 "Spacetime transitions in string theory", IAS preprint IASSNS 93/36.
\item C.M. Hull and B. Spence, Phys. Lett. {\bf 232B}, 204(1989).
\item C. Emmrich and H. R\"omer, Commun. Math. Phys. {\bf 129}, 69(1990).
\item L. Dixon, J.A. Harvey, C. Vafa and E. Witten, Nucl. Phys. {\bf B261},
678(1985); {\bf B274}, 285(1986).
\item D. Mumford, Algebraic Geometry I, Complex Projective Varieties,
Springer-Verlag, Berlin, (1976)\\
R. Hartshorne, Algebraic Geometry, Springer-Verlag, New York, (1977).
\item P. Candelas and X. C. de la Ossa, Nucl. Phys. {\bf B342}, 242(1990),\\
P. Candelas, X.C. de la Ossa, P.S. Green and L. Parkes, Nucl. Phys.
{\bf B359}, 21(1991).
\item B.R. Greene, S.-S. Roan and S.-T. Yau, Commun. Math. Phys. {\bf 142},
245(1991).
\item M. Gromov, Partial Differential Relations, Springer-Verlag,
 Berlin-Heidelberg-New York, (1986), p 340.
\item V. Guillemin and S. Sternberg, Invent. Math. {\bf 97}, 485(1989).
\item M. Audin, The Topology of Torus Actions on Symplectic Manifolds,
Progress in Mathematics Vol. 93, Birkh\"auser-Verlag, Basel, (1991).
\item J.J. Duistermaat and G. Heckman, Invent. Math. {\bf 69}, 259(1982).
\item M. Atiyah, Bull. Lond. Math. Soc. {\bf 14}, 1(1982);\\
V. Guillemin and S. Sternberg, Invent. Math. {\bf 67}, 491(1982).
\item See e.g., S. Helgason, Differential Geometry and Symmetric Spaces,
Acad. Press, New York, (1962)\\
S. Kobayashi and K. Nomizu, Foundations of Differential Geometry, Vol. I,
Wiley-Interscience, New York, (1963)
\item V. Guillemin and S. Sternberg, Symplectic Techniques in Physics,
Cambridge Univ. Press, Cambridge, (1984).
\item S.S. Chern, Complex Manifolds without Potential Theory, 2nd Edition,
 Springer-Verlag, New York, (1979).
\item E. Witten, Nucl. Phys. {\bf B403}, 185(1993).
\item A. Floer, Commun. Math. Phys. {\bf 120}, 575(1989).
\item E. Witten, Commun. Math. Phys. {\bf 118}, 411(1988).
\item J. Milnor, Singular Points of Complex Hypersurfaces, Princeton Univ.
Press, Princeton, N. J. (1968).
\item V.I. Arnold, S.M. Gusein-Zade and A.N. Varchenko, Singularities of
Differentiable Maps, Vol. II, Birkh\"auser, Boston, (1988), p29.
\item L. P. Eisenhart, Riemannian Geometry, Princeton Univ. Press, Princeton,
N.J. (1965).
\item J. Honerkamp, Nucl. Phys. {\bf B36}, 130(1972).
\item L. Alvarez-Gaume and D.Z. Freedman, Annals of Phys. {\bf 134}, 85(1981).
\item R. Sjamaar and E. Lerman, Ann. Math. {\bf 134}, 375(1991).
\item A.M. Polyakov, Gauge Fields and Strings, Harwood Academic Publishers,
 Chur (1987), p142.
\item P.S. Aspinwall, B.R. Greene and D.R. Morrison, preprints IASSNS-HEP
93/38, CLNS-93/1236; IASSNS-HEP 93/49.
\item D. McDuff, Topology {\bf 30}, 409(1991); J. Geom. Phys. {\bf 5},
149(1988).
\item W. Hurewicz, Lectures in Ordinary Differential Equations, M.I.T. Press,
 Cambridge, Mass. (1966).
\end{enumerate}
\end{document}